# Interaction of Nonlinear Schrödinger Solitons with an External Potential

Helge Frauenkron and Peter Grassberger

*Physics Department, University of Wuppertal,*
*D–42097 Wuppertal, Germany*

June 28, 1995

**Abstract**

Employing a particularly suitable higher order symplectic integration algorithm, we integrate the 1-$d$ nonlinear Schrödinger equation numerically for solitons moving in external potentials. In particular, we study the scattering off an interface separating two regions of constant potential. We find that the soliton can break up into two solitons, eventually accompanied by radiation of non-solitary waves. Reflection coefficients and inelasticities are computed as functions of the height of the potential step and of its steepness.

# 1 Introduction

Recent years have seen a considerable growth in the interest for nonlinear partial differential equations with soliton solutions. In particular, the nonlinear Schrödinger equation (NLSE) and its variants appear in problems drawn from disciplines as diverse as optics, solid state, particle and plasma physics. There, the NLSE describes phenomena such as modulational instability of water waves [1], propagation of heat pulses in anharmonic crystals, helical motion of very thin vortex filaments, nonlinear modulation of collisionless plasma waves [2], and self-trapping of light beams in optically nonlinear media [3, 4, 5]. In all these problems, the main interest is in the fact that the NLSE has soliton solutions. These are solitary waves with well defined pulse-like shapes and remarkable stability properties [6].

A great deal of current interest is directed to the question how these states behave under the influence of external perturbations. These can be of various forms. We shall limit ourselves to such perturbations which can be described by potentials. They preserve the hamiltonian structure of the NLSE [2], but not its complete integrability. Other types of perturbation which also are hamiltonian are obtained when either the coefficient in the kinetic term (the 'mass' in the quantum mechanical interpretation) or in the nonlinear term are made spatially not constant. Such inhomogeneities have indeed been studied more intensely than the ones we shall study below, since they are more relevant for the transmission of pulses through junctions in optical fibers [4, 5, 7, 8, 9].

More precisely, we shall consider only potentials which are constant outside a finite interval (we shall only consider the case of one spatial dimension). But we allow different values $V_\pm$ for $x \to \pm\infty$, mimicking thereby the effect of an interface between two media in which the solitons have different characteristics. We study initial conditions consisting of one single soliton. In general, we have to expect that this soliton will not just be transmitted or reflected. There might be also inelastic scatterings where it breaks up either into several solitons or into non-solitary waves, or both.

This problem has been studied previously by several authors. While perturbative approaches were used in [3, 4, 5, 10], straightforward numerical integrations were made in [11]. Both approaches showed that the soliton behaves just like a classical particle if the force created by the potential is sufficiently weak. This is to be expected, but the problem what happens when the force is strong was left open for a simple potential ramp (the potential considered in [5] was more complicated).

It is one purpose of the present paper to close this gap by means of simulations.

Another purpose is to show the usefulness of higher order symplectic integration algorithms. As we have already mentioned, the NLSE is a hamiltonian system. Thus, it is natural to apply to it integration routines which were developed during the recent years and whose main characteristic is that they preserve the hamiltonian structure [12, 13, 14]. The latter is not true e.g. for standard methods as e.g. Runge-Kutta or predictor-corrector. Such 'symplectic' integrators (the simplest of which is the well known Verlet or 'leap frog' algorithm) have been applied already to the linear [15, 16, 17, 18] and nonlinear [19, 20, 21, 22] Schrödinger equations.

The most popular algorithms of this type are split-operator methods. They depend on the hamiltonian being a sum of two terms $A$ and $B$, each of which can be integrated explicitly. Then one uses the Baker-Campbell-Hausdorff theorem to approximate $e^{i(A+B)t}$ by a product of factors $e^{i\alpha_k At}$ and $e^{i\beta_k Bt}$, where $\alpha_k$ and $\beta_k$ are real numbers satisfying among others $\sum_k \alpha_k = \sum_k \beta_k = 1$. The error is then given by higher order commutators of $A$ and $B$. We shall in particular apply a fourth order method due to McLachlan and Atela [23] which is applicable if one of the third order commutators vanished identically. We shall see that this method should be applicable to our problem, and that it is indeed numerically very precise, indicating that the McLachlan-Atela method is the method of choice for a wide class of problems.

## 2 The NLSE soliton solution

Using appropriate units, we can write the NLSE as

$$i\frac{\partial\Psi(x,t)}{\partial t} = -\frac{1}{2}\frac{\partial^2\Psi(x,t)}{\partial x^2} - |\Psi(x,t)|^2\Psi(x,t) + V(x)\ \Psi(x,t), \tag{1}$$

where $V(x)$ is the external potential. We shall use for the latter a piecewise linear ansatz, with $V(x) \equiv 0$ for $x < 0$, $V(x) \equiv V_0 > 0$ for $x > x_0 \geq 0$, and linearly rising for $x$ between 0 and $x_0$,

$$V(x) = \begin{cases} 0 & : \quad x < 0 \\ xV_0/x_0 & : \quad 0 \leq x < x_0 \\ V_0 & : \quad x \geq x_0 \end{cases} . \tag{2}$$

We call the negative $x$-axis region $I$, while region $II$ is the region $x > 0$ (where $V(x) > 0$).

We study scattering solutions where the incoming wave consists of a single soliton arriving from region $I$. The outgoing wave will then in general be a complicated superposition of solitons and non-solitary waves, in general moving both into regions $I$ and $II$. The interesting questions are how many solitons will leave the scattering region and with what energies, how much of the total energy is transmitted and reflected, and how much of it goes into non-solitary waves.

For a constant potential $V_0$ the soliton solutions of eq.(1) form a two-parameter manifold (apart from translations). Taking as parameters the velocity $v$ and the amplitude $a$, these solutions read [24]

$$\Psi(x,t) = \frac{a}{\cosh[a(x-vt)]}e^{i\{vx+[(a^2-v^2)/2-V_0]t\}} . \tag{3}$$

We denote the velocity of the incoming soliton as $v_0$. Using a suitable rescaling of $x, t$ and $\Psi$, we can always choose its amplitude as $a_0 = 1/2$, without loss of generality.

Among the infinitely many conserved quantities (for $V(x) = \text{const!}$) the following three are of particular interest:
the normalization

$$N = \int |\Psi|^2\ dx\ , \tag{4}$$

the energy

$$E = \int \left( \frac{1}{2}\left|\frac{\partial\Psi}{\partial x}\right|^2 - \frac{1}{2}|\Psi|^4 + V(x)\ |\Psi|^2 \right)\ dx\ , \tag{5}$$

and the momentum

$$P = \frac{1}{2i}\int \left( \Psi^*\frac{\partial\Psi}{\partial x} - \Psi\frac{\partial\Psi^*}{\partial x} \right)\ dx\ . \tag{6}$$

For the soliton given by eq.(3), $N = 2a$, $P = vN$, and $E = (v^2/2 - a^2/6)N + \langle V\rangle N$, where the average over $V(x)$ is taken with weight $\propto |\Psi|^2$ as indicated by eq.(5).

For a slowly varying $V(x)$ (which implies $x_0/V_0 \gg 1$ in our case) the amplitude is approximately constant, and the soliton moves like a classical particle with mass $m = 2a$ in an external potential $mV(x)$ [10]. The mass of the incoming soliton is $m_0 = 1$ with our normalization. Another limit case where the soliton behaves like a particle is that of $V_0 \ll K_0$ where $K_0 = v_0^2/2$ is the kinetic energy of the incoming soliton.

It is easily seen that $N$ and $E$ are also conserved for non-constant potential $V$, while this is not true for $P$. Denoting by $N_i$, $i = I, II$, the normalization in region $i$, we have thus $N_{I,\text{out}} + N_{II,\text{out}} = N_{I,\text{in}} = 1$. Similarly, energy conservation gives $E_{I,\text{out}} + E_{II,\text{out}} = E_{I,\text{in}} = (v_0^2 - 1/12)/2$.

Conservation of $N$ and $E$ poses restrictions on the final state. In general, they do not seem to be very stringent. Assume e.g. that the final state consists of two solitons moving in opposite directions, $(a, v)$ moving into region $I$ and $(b, w)$ moving into $II$. Then we find that

$$a + b = 1/2, \qquad v_0^2 = ab + 2(av^2 + bw^2 + 2bV_0) \tag{7}$$

This does not imply, in particular, a lower bound on $v_0$ since $b$ and $v$ can be arbitrarily small. Similarly, for any initial soliton we can have any number of outgoing solitons, provided there is at least one reflected and one transmitted soliton. Conservation of $N$ and $E$ is more stringent if no or all solitons are reflected. For instance, if the final state consists of a single transmitted soliton, then its velocity is $v_{II,\text{out}} = \sqrt{v_0^2 - 2V_0}$. This conforms with the general statement that the soliton behaves like a classical particle with $m = 1$, and shows that there is no transmission if $v_0 < \sqrt{2V_0}$ (i.e., $K_0 < V_0$) and $x_0 \gg V_0$. It was verified numerically in [11]. These authors concluded indeed that solitons impinging on a potential step behave like classical particles. It was mainly this claim which stimulated our investigation.

# 3 Symplectic integration

The NLSE is a classical hamiltonian system with Poisson bracket

$$\{\Psi^*(x), \Psi(y)\} = i\delta(x - y) \tag{8}$$

and hamiltonian $H = E$. This implies in particular that it can be written as

$$\dot{\Psi} = \{\Psi, H\} = \mathcal{H}\Psi\,, \tag{9}$$

where the linear ('Liouville') operator $\mathcal{H}$ is defined as $\mathcal{H}\,\cdot = \{\,\cdot\,, H\}$. Split-operator methods can be applied by splitting $\mathcal{H} = \mathcal{T} + \mathcal{V}$, where $\mathcal{T}$ and $\mathcal{V}$ are the Liouvilleans corresponding to $\frac{1}{2}\int dx|\partial_x\Psi|^2$ and $\int dx(-\frac{1}{2}|\Psi|^4 + V|\Psi|^2)$,

$$\mathcal{T}\Psi = \frac{i}{2}\partial_x^2\Psi\ , \quad \mathcal{V}\Psi = i(|\Psi|^2\Psi - V\Psi)\ . \tag{10}$$

In a paper by McLachlan & Atela [23], a fourth order algorithm was introduced which minimizes the neglected fifth order terms in the Baker-Campbell-Hausdorff formula for hamiltonians for which

$$\{\{\{\mathcal{T},\mathcal{V}\},\mathcal{V}\},\mathcal{V}\} \equiv 0 \ . \tag{11}$$

This applies obviously to hamiltonians with $T = \frac{1}{2}(p, M^{-1}p)$, $V = V(x)$, with $M$ a constant mass matrix and $\{q_i, p_k\} = \delta_{ik}$, since there each commutator with $V$ acts as a derivative operator on any function of $p$. In [17] it was shown that this algorithm can also be applied to the linear SE where it gives better performance than the general fourth order algorithm [12] which does not take into account this special structure.

Although the argument is less straightforward in the present case, it is not too hard to see that eq.(11) holds also there [22]. Let $f(|\Psi|^2, x)$ and $g(|\Psi|^2, x)$ be arbitrary functions with finite first and second derivatives. Then one finds

$$\Big\{\int dx |\partial_x \Psi|^2 g(|\Psi|^2, x), \int dy f(|\Psi|^2, y)\Big\} = i \int dx (\Psi^*_{xx}\Psi - \Psi^*\Psi_{xx}) g f' \ , \tag{12}$$

where $f' = \partial f / \partial |\Psi^2|$, and

$$\Big\{\int dx (\Psi^*_{xx}\Psi - \Psi^*\Psi_{xx}) g, \int dy f\Big\} = -2i \int dx |\Psi|^2 \frac{dg(|\Psi(x)|^2, x)}{dx} \frac{df'(|\Psi(x)|^2, x)}{dx} \ . \tag{13}$$

Since the last expression is a functional of $|\Psi|^2$ only, its Poisson bracket with $\int dy f$ vanishes identically, QED.

The coefficients $\alpha_k$ and $\beta_k$ for the McLachlan-Atela method are listed in [23, 17]. Our implementation involves a spatial grid with Fourier transformation after each half step [17].

Since $\mathcal{T}$ and $\mathcal{V}$ both conserve the normalization exactly, $N$ should be conserved up to round-off errors. This was checked numerically, relative errors typically were of order $10^{-11}$. Energy is not conserved exactly, and its error was $\approx 10^{-5}$ after an evolution time $t = 300$ with an integration step $\Delta t = 0.005$. The precise value depended of course on the parameters of the soliton and on $x_0$. It was checked that the algorithm is indeed fourth order, and is more precise than the general fourth order symplectic [12] and the leap-frog (second order symplectic) algorithms. We also tested two other discrete Hamiltonian integration schemes which where examined in [26]. They both show the same qualitative behavior, but the discretization of the Laplace operator requires smaller time steps for the same spatial discretization width. All this demonstrates the advantage of the McLachlan-Atela algorithm.

## 4 Results

During the simulations we measured normalization $N_i$, energy $E_i$, and momentum $P_i$ in each region ($i = I, II$) separately. The derivatives of $\Psi$ and $\Psi^*$ were of course computed in Fourier space, as this is much more precise than taking finite differences in $x$-space.

Since we have two conserved quantities, we can define two sets of transmission and reflection coefficients. We call them $T_N, R_N$ and $T_E, R_E$,

$$T_N = \frac{N_{II}}{N} , \qquad R_N = \frac{N_I}{N} = 1 - T_N \tag{14}$$

and

$$T_E = \frac{E_{II}}{E} , \qquad R_E = \frac{E_I}{E} = 1 - T_E \ . \tag{15}$$

In addition we registered all local maxima of $|\Psi(x)|^2$ with $|\Psi(x)|^2 > 1/3000$.

Since our model involves 3 free parameters $(V_0, x_0, v_0)$, it is impossible to present results exhaustively. We did a large number of simulations with different parameter values, but we present only a few of them here to illustrate the variety of the scenarios.

Our numerical simulations confirmed the prediction that the soliton behaves as a classical particle if $x_0/V_0 \gg 1$, and if $V_0 \ll K_0$. The same is true also if $x_0 = 0$ and $V_0 = \infty$, i.e. if the potential acts like a hard wall. In that case, an exact solution of the NLSE with boundary condition $\Psi|_{x=0} = 0$ and correct initial conditions in region $I$ is provided by a state with two (interacting) solitons with opposite velocities and phases but equal amplitudes [11].

While the above essentially just checked the correctness of our integration routine, a less trivial result is that we confirmed the observations of Nogami and Toyama [11] for their parameter choice $x_0 = 0, v_0 = 0.2, V_0 \approx K_0$. But we did *not* verify their claim that this is the typical behavior. Instead, the soliton typically breaks up and does not behave like a classical particle.

In general, after the soliton hits the potential ramp, we found typically more than a single maximum of $|\Psi(x)|$. Moreover, the heights of these maxima in general were not constant in time, though they moved with practically constant velocities (see figs. 1, 3, 5, 7). Instead, they showed often very marked oscillations (fig. 2, 4, 6, 8) which were damped in all cases. Such damped oscillations result typically from superpositions of solitons with non-solitary waves [27]. We checked that a superposition of a soliton with a Gaussian wave packet gave essentially the same patterns.

In the following we shall only show results for $v_0 = 0.8$ although, as we said, we had made runs also with different $v_0$ and with similar results in general.

Figures 1 and 2 show the case where the potential is a step function ($x_0 = 0$) and the kinetic energy ($K_0 = 0.32$) is larger than its height $V_0 = 0.3$. Classically one would expect the soliton to move into region $I$ and to propagate there with a reduced speed. But our simulation shows that it breaks up into two solitons with roughly equal heights and with velocities $v = -0.588$ and $w = 0.395$. About half of the normalization and three thirds of the energy are transmitted ($T_N = 0.527, T_E = 0.712$). Inserting these numbers into eq.(7), we find perfect agreement (discrepancies are $\lesssim 1\%$). This indicates that radiation in form of non-solitary waves is small in spite of the wiggles seen in fig. 2. More precisely, we compared our data with ref.[27] by assuming that the transmitted wave is a single solitary wave immediately after leaving the interaction region. We found perfect agreement if we assume that this wave has exactly the same shape and width as the incoming soliton, but an amplitude reduced by a factor 0.728. Thus, at least for these parameter values, the main effect of the interaction on the transmitted wave is simply a reduction of amplitude.

The situation where the potential step ($V_0 = 0.34, x_0 = 0$) is higher than the kinetic energy $K_0$ is plotted in the figures 3 and 4. Here one would expect classically that the incident soliton is completely reflected back into region $II$. But once again the behavior is quite different, the soliton splits up into two. The transmitted one is not as high as the reflected one ($T_N = 0.373$) and therefore much wider, but it still carries more than half of the initial energy, $T_E = 0.571$. As we increase $V_0$ further, the transmitted soliton rapidly shrinks. It becomes unobservable at $V_0 \approx 2K_0$, where the soliton is practically completely reflected.

Let us now study positive values of $x_0$, i.e. potential ramps with finite slopes. Our data show unambiguously that this slope has a strong influence. If $x_0$ is of order 1, the soliton still breaks up as described above (figs. 5, 6), with even larger oscillations and even more "dirt" than for $x_0 = 0$. Flattening the potential ramp further but leaving its height constant, the soliton finally travels along the classically expected trajectory (figs. 7, 8): in the ramp region it sees a constant force and hence moves on a parabola; it is reflected (transmitted) for $V_0 > K_0$ ($V_0 < K_0$).

This dependence on the slope of the ramp is seen very clearly when plotting the energy in region $II$ as a function of time, see fig. 9. While the asymptotic state is reached very quickly for steep potentials, this evolution takes very long for gentle slopes. If $x_0 \gg 1$ (corresponding to a width of the soliton $\ll x_0$), the energy change is sudden when the soliton crosses the point $x = 0$.

Finally, the dependence of the transmission coefficients on $x_0$ are shown in fig. 10. We see that they are not monotonic, with the nonmonotonicity more pronounced for $T_N$ than for $T_E$. This is an unexpected effect which we do not know how to explain. The fact that $T_N < T_E$ for all $x_0$ is less surprising.

# 5 Summary and conclusions

In this note we have applied an optimized fourth order symplectic integrator to the scattering of NLSE solitons from an external potential. The integrator is optimized in the sense that it takes into account that the kinetic energy is bilinear in $\Psi_x$. It was found to be more precise than the general fourth order symplectic integrator. We found that solitons break in general up when hitting a potential threshold, in contrast to recent claims. The complexity of the outgoing state depends on the parameters of the potential and of the soliton, but most frequently the soliton breaks into two, with rather little radiation.

The NLSE can be considered as a special case of the complex Ginzburg-Landau (CGL) equation $\dot{\Phi} = \mu\Phi + \alpha|\Phi|^2\Phi + \beta\nabla^2\Phi$ ($\mu, \alpha, \beta \in \mathbb{C}$) with complex constants. The applicability of our integrator does not depend on the phases of these terms, whence it should be applicable also to the CLG equation in general. We just have to take into account that $|\Phi|$ is not constant dusssring the evolution under the nonlinear term if $\mathrm{Re}\,\mu, \alpha \neq 0$. In that case the integration of $\mathcal{V}$ involves solving the easy differential equation $d|\Phi|^2/dt = 2(\mathrm{Re}\,\mu\,|\Phi|^2 + \mathrm{Re}\,\alpha\,|\Phi|^4)$.

This work was partly supported by DFG within the Graduiertenkolleg "Feldtheoretische und numerische Methoden in der Elementarteilchen- und Statistischen Physik", and within SFB 237.

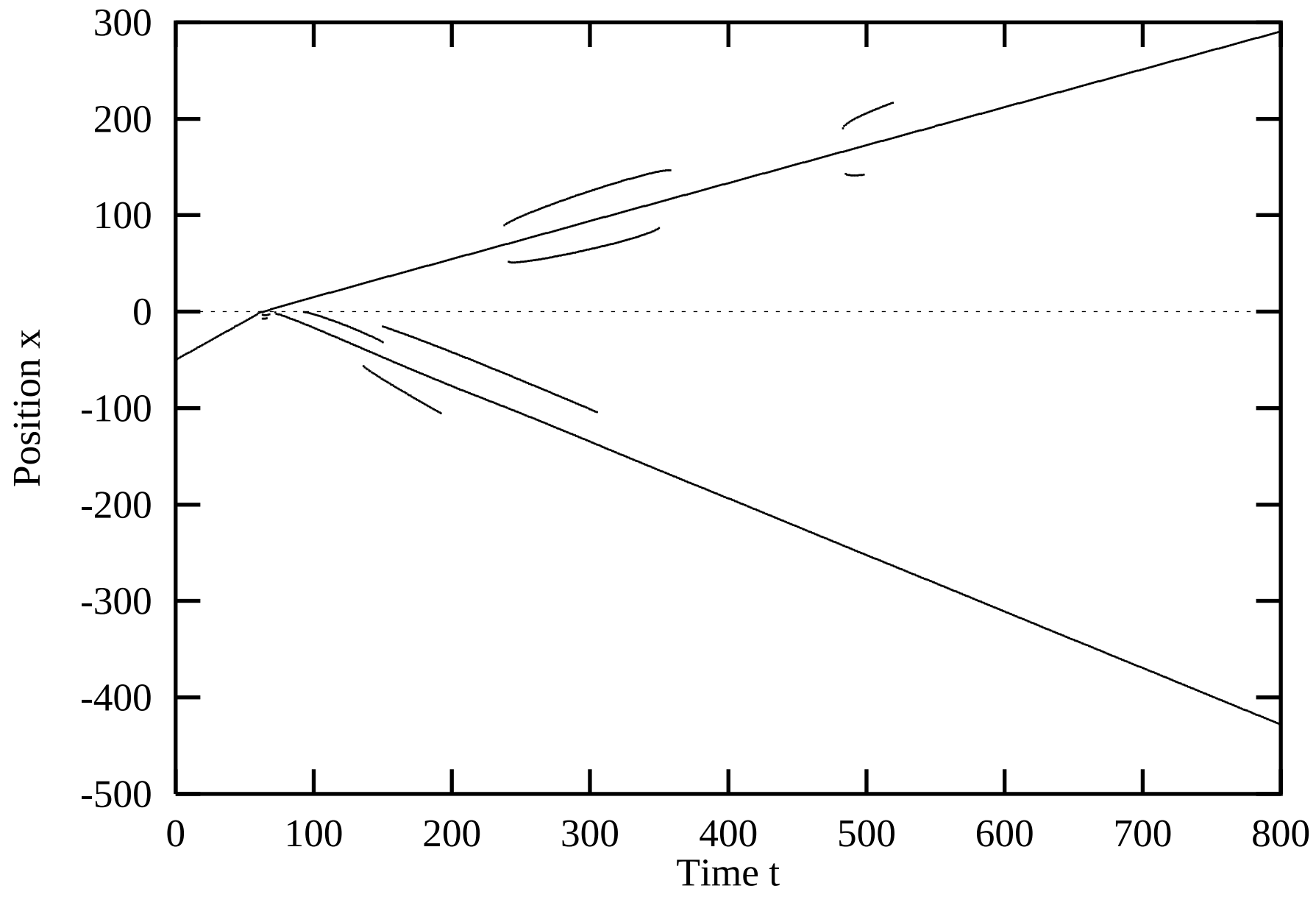

Figure 1: Time evolution of local maxima of $|\Psi|^2$ for a soliton with incident velocity $v_0 = 0.8$ which is scattered at a potential step with $x_0 = 0$ and height $V_0 = 0.3 = 0.937K_0$. The calculation was done on a lattice with 4096 sites, discretization width $\Delta x = 0.2$ and integration step $\Delta t = 0.005$. The latter parameters are the same for the next figures.

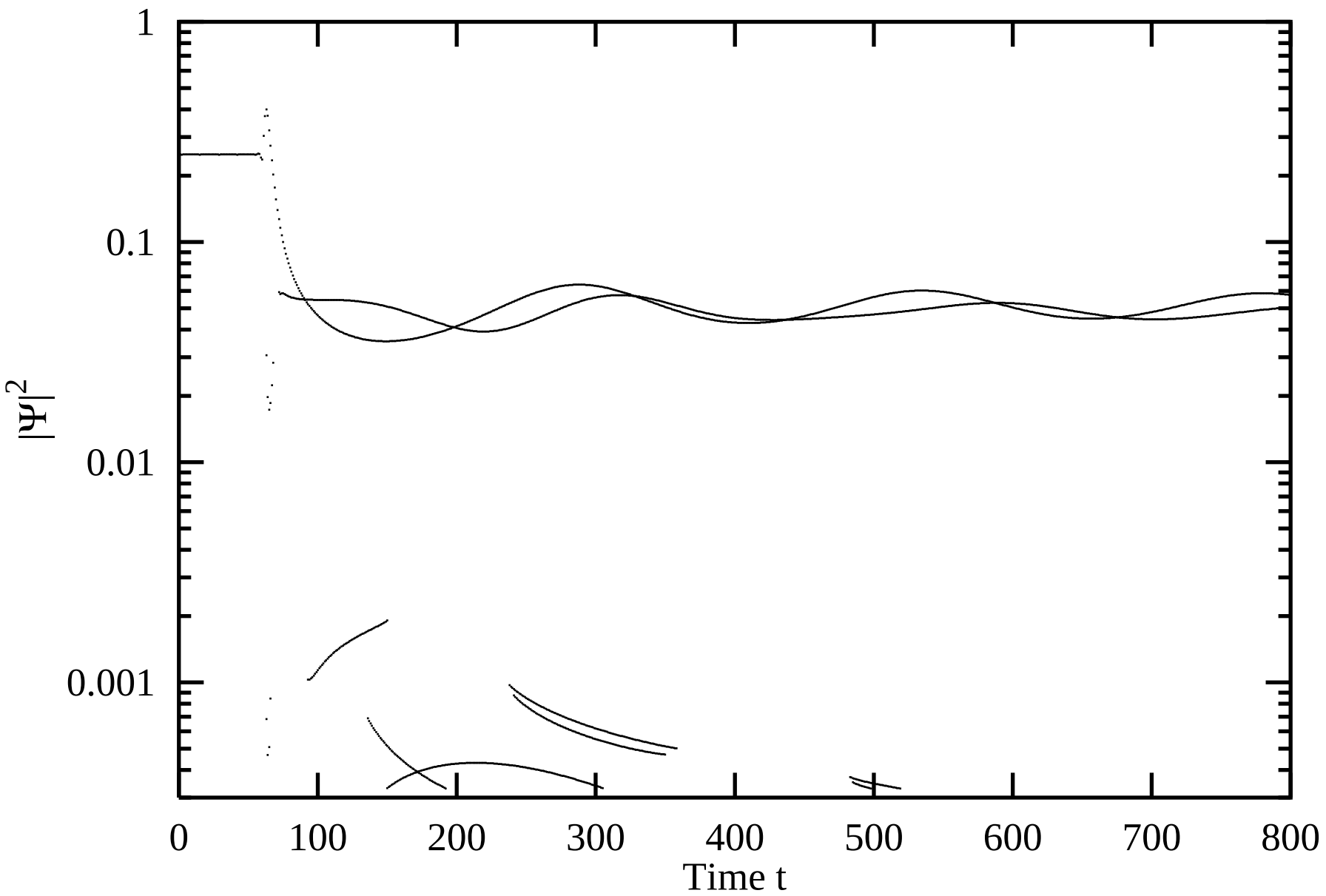

Figure 2: Time evolution of the height of the maxima shown in fig. 1. The highest curve belongs to the transmitted soliton, and the second highest to the reflected one. The other maxima presumably are due to the superposition of non-solitary waves.

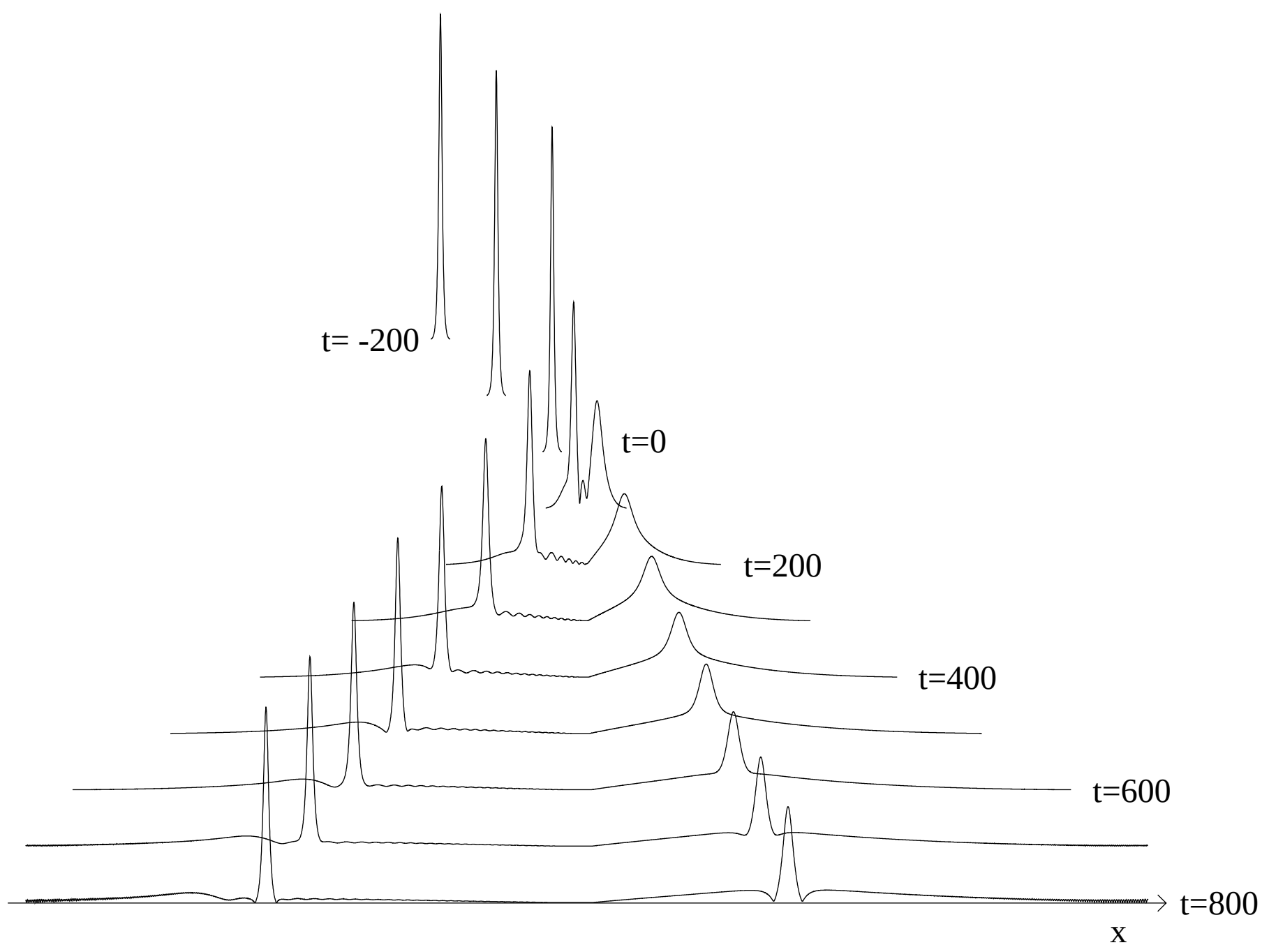

Figure 3: Time evolution of $|\Psi|^2$ shown in a 3-dimensional plot with $V_0 = 0.34 = 1.063 K_0$.

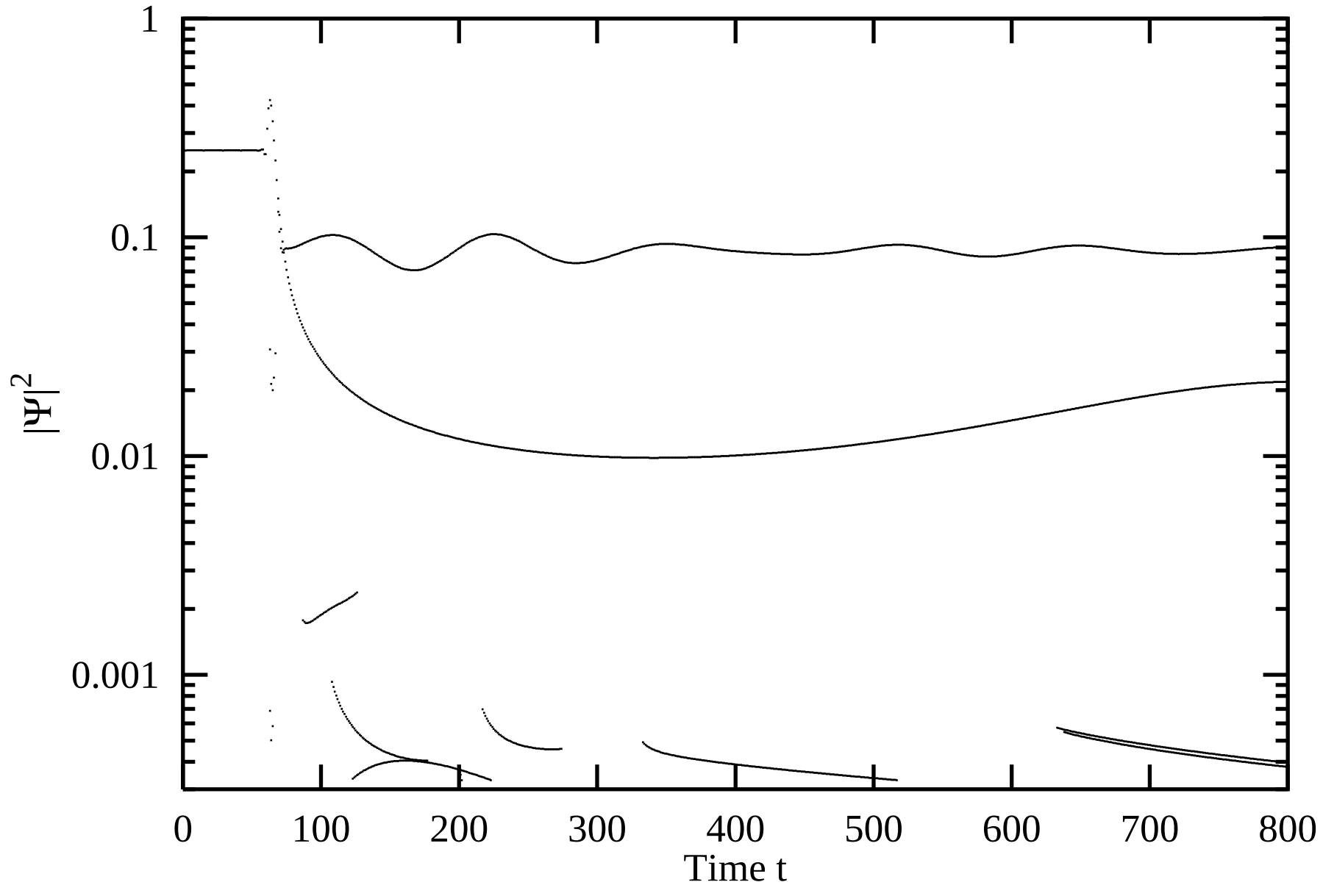

Figure 4: Same as fig.2, but for $V_0 = 0.34$ as in fig.3. Now the highest curve belongs to the reflected soliton.

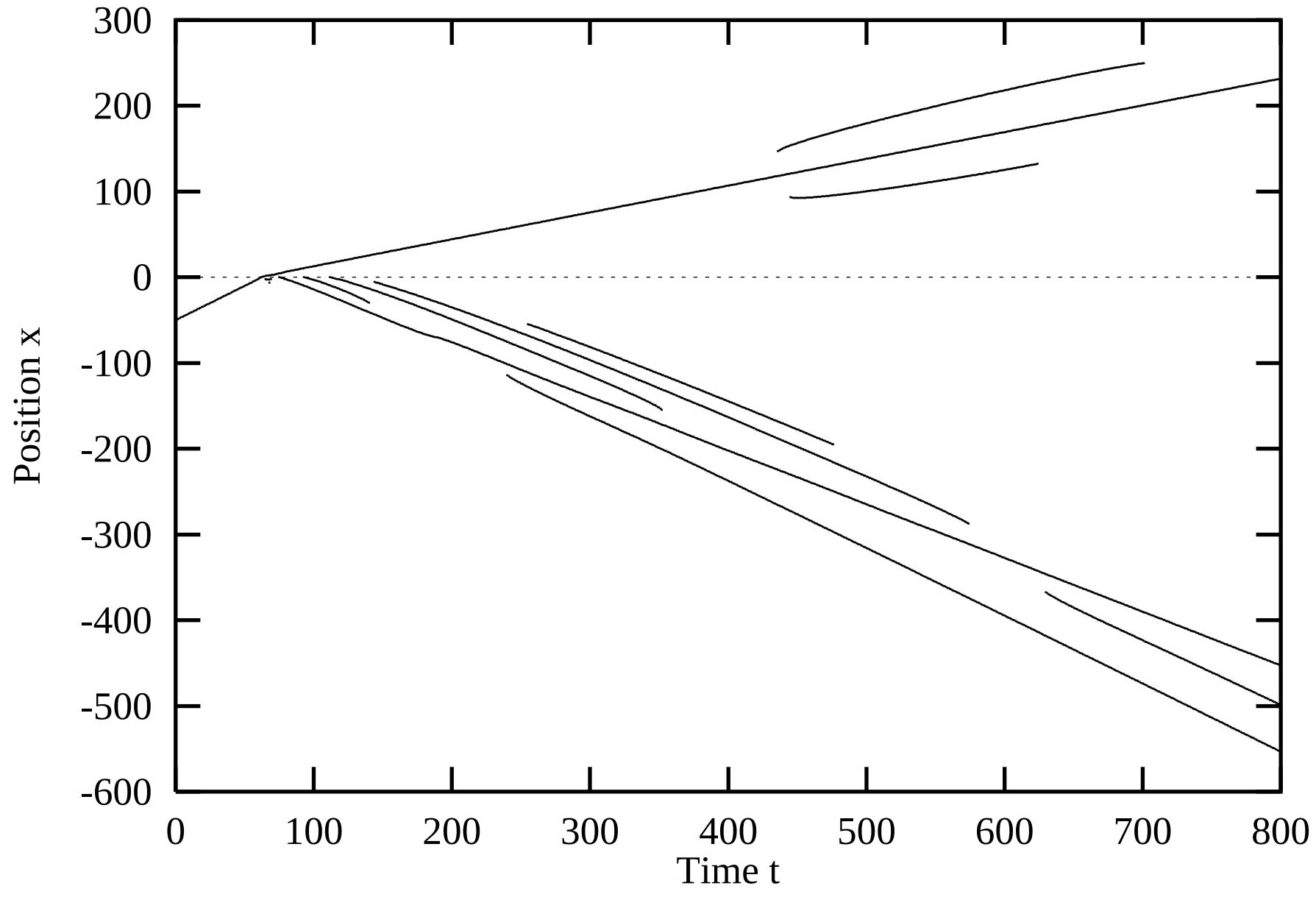

Figure 5: Same as fig.1, but for $x_0 = 3$ and $V_0 = 0.35 = 1.094 K_0$.

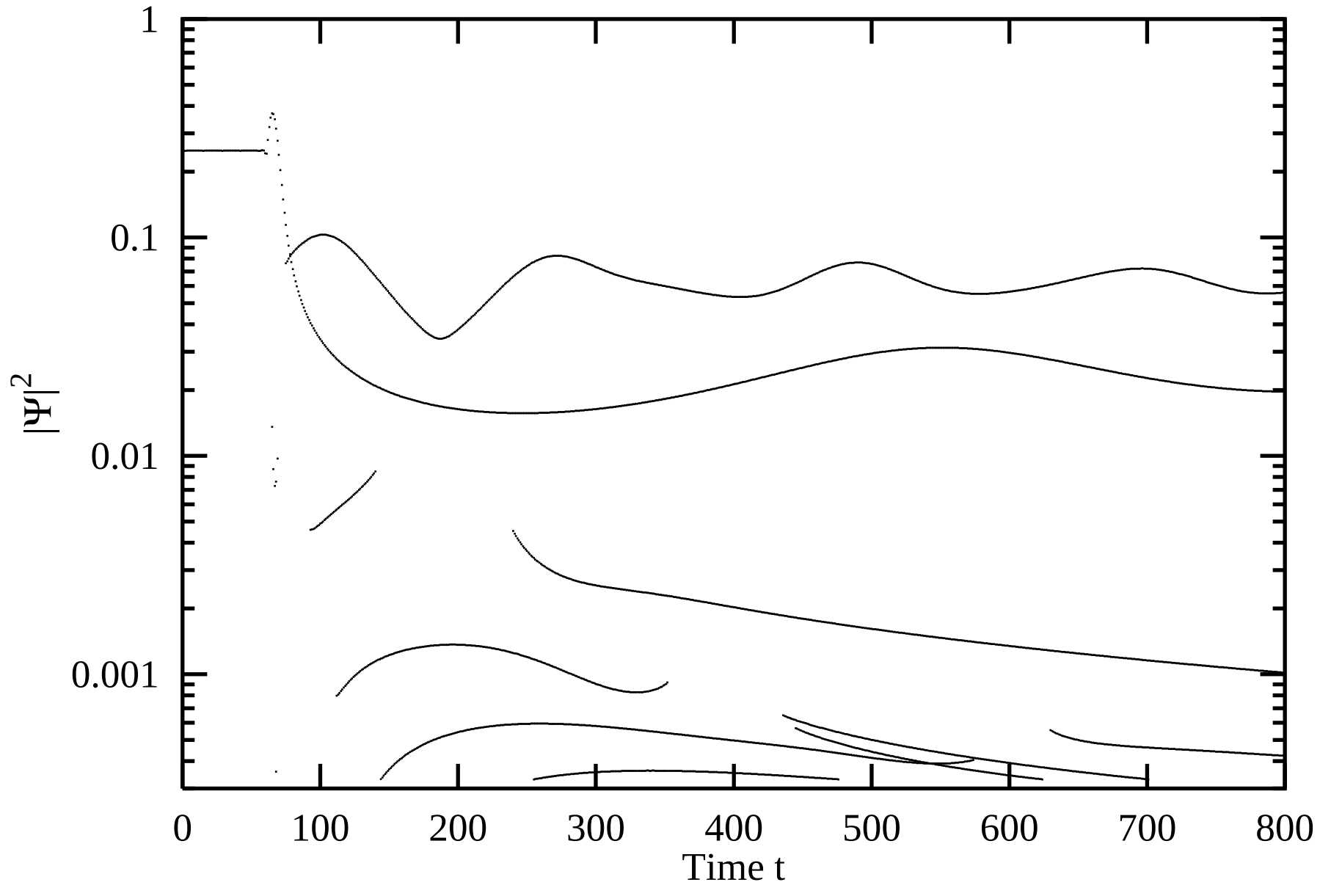

Figure 6: Same as fig.2, but with $x_0$ and $V_0$ as in fig.5. The highest curve belongs to the reflected soliton and the second highest to the transmitted one. The other maxima are side maxima presumably due to the superposition of non-solitary waves.

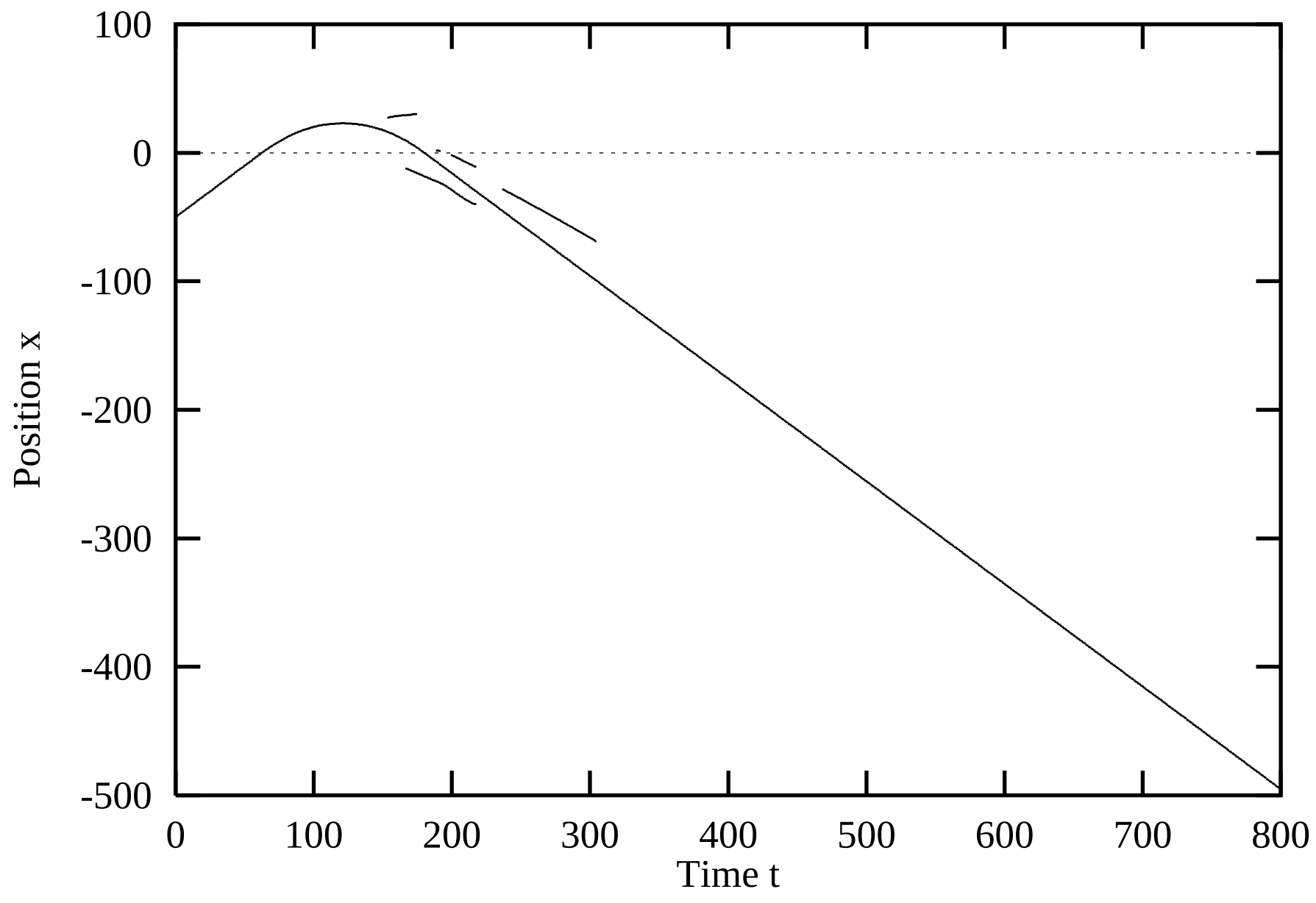

Figure 7: Same as fig.1, but for $x_0 = 25$ and $V_0 = 0.35 = 1.094K_0$.

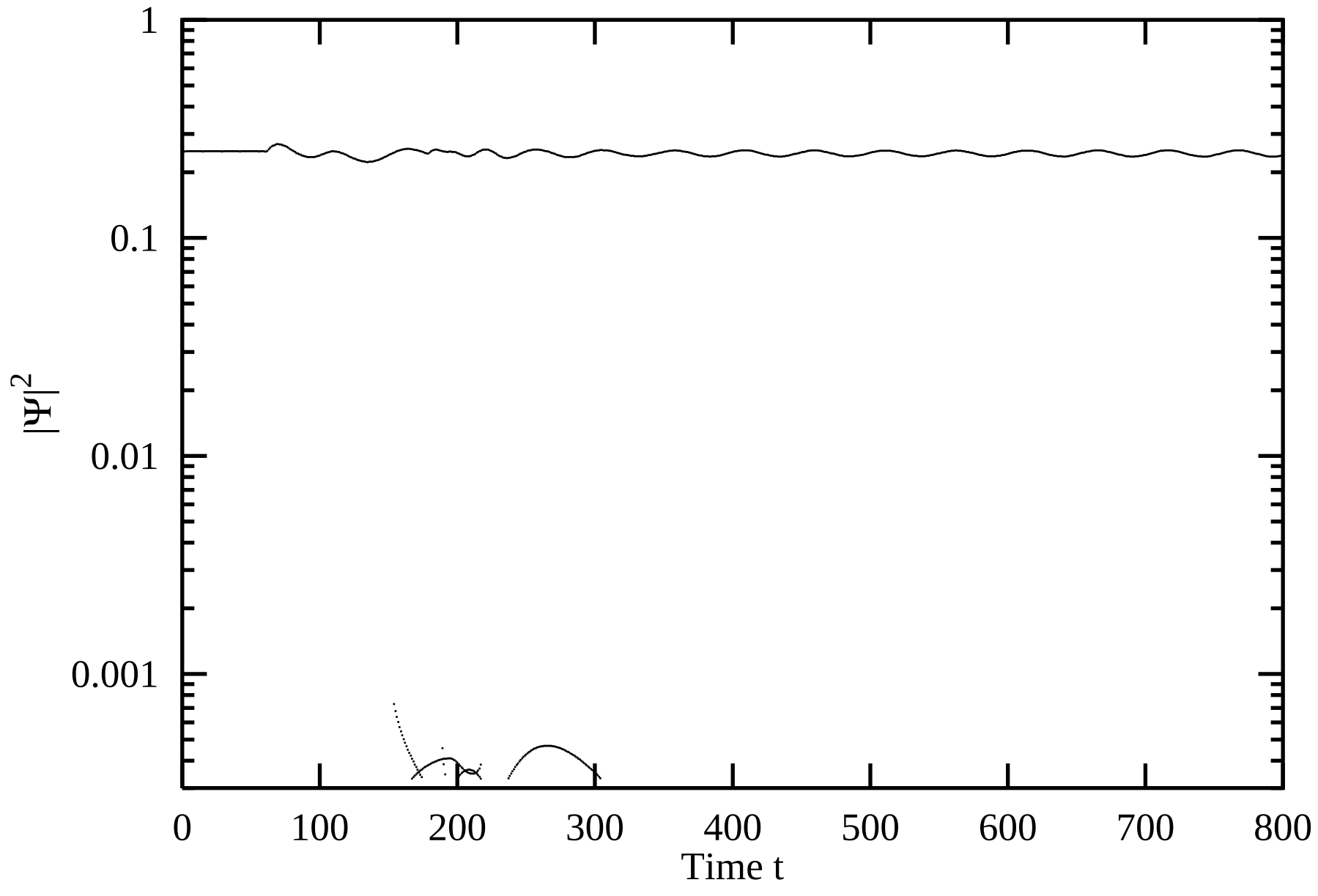

Figure 8: Same as fig.2, but with $x_0$ and $V_0$ as in fig.5.

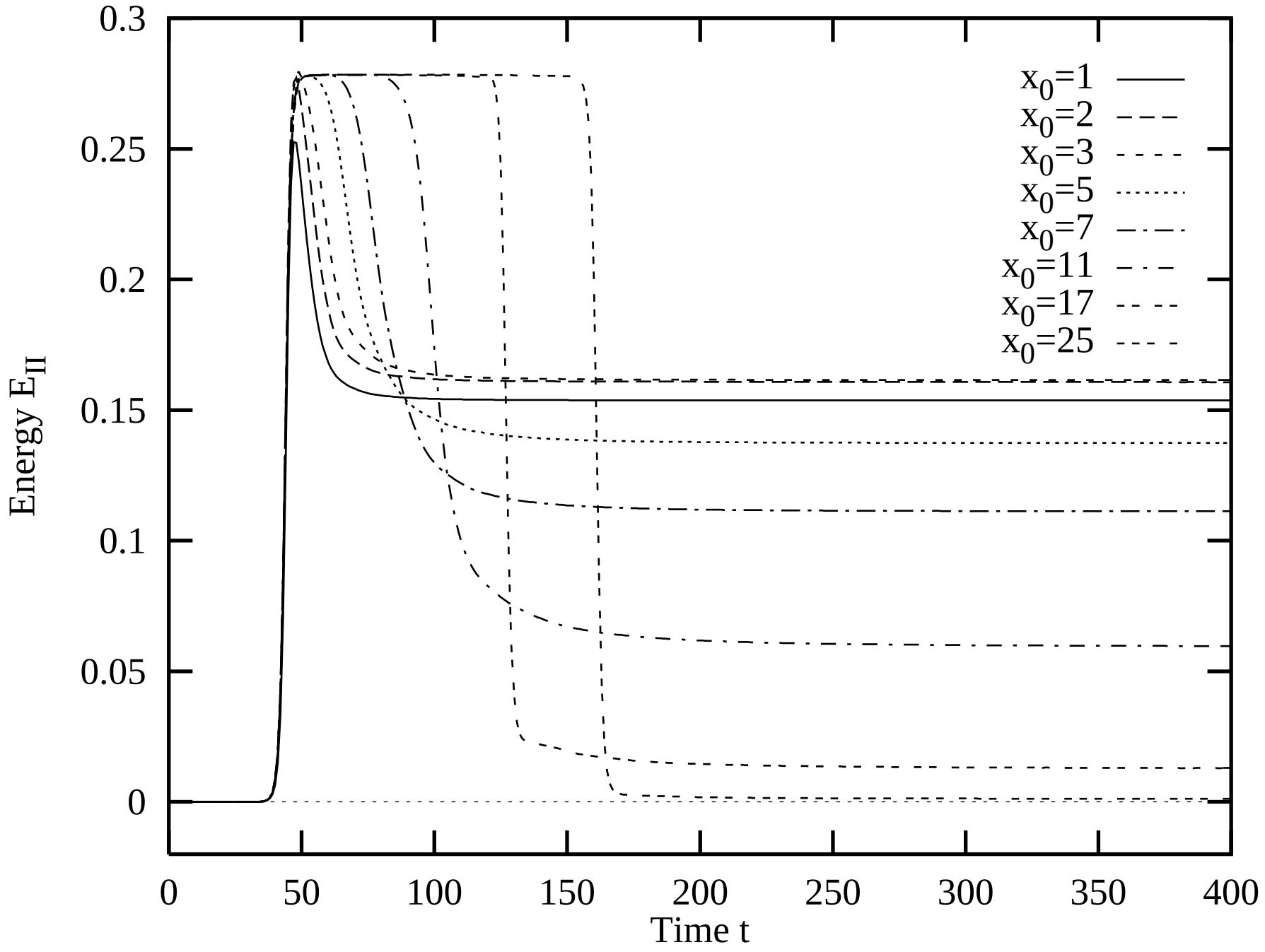

Figure 9: Time evolution of $E_{II}$, i.e. the energy in region $II$, for different values of $x_0$. For all curves, $v_0 = 0.8$ and $V_0 = 0.35 = 1.094K_0$.

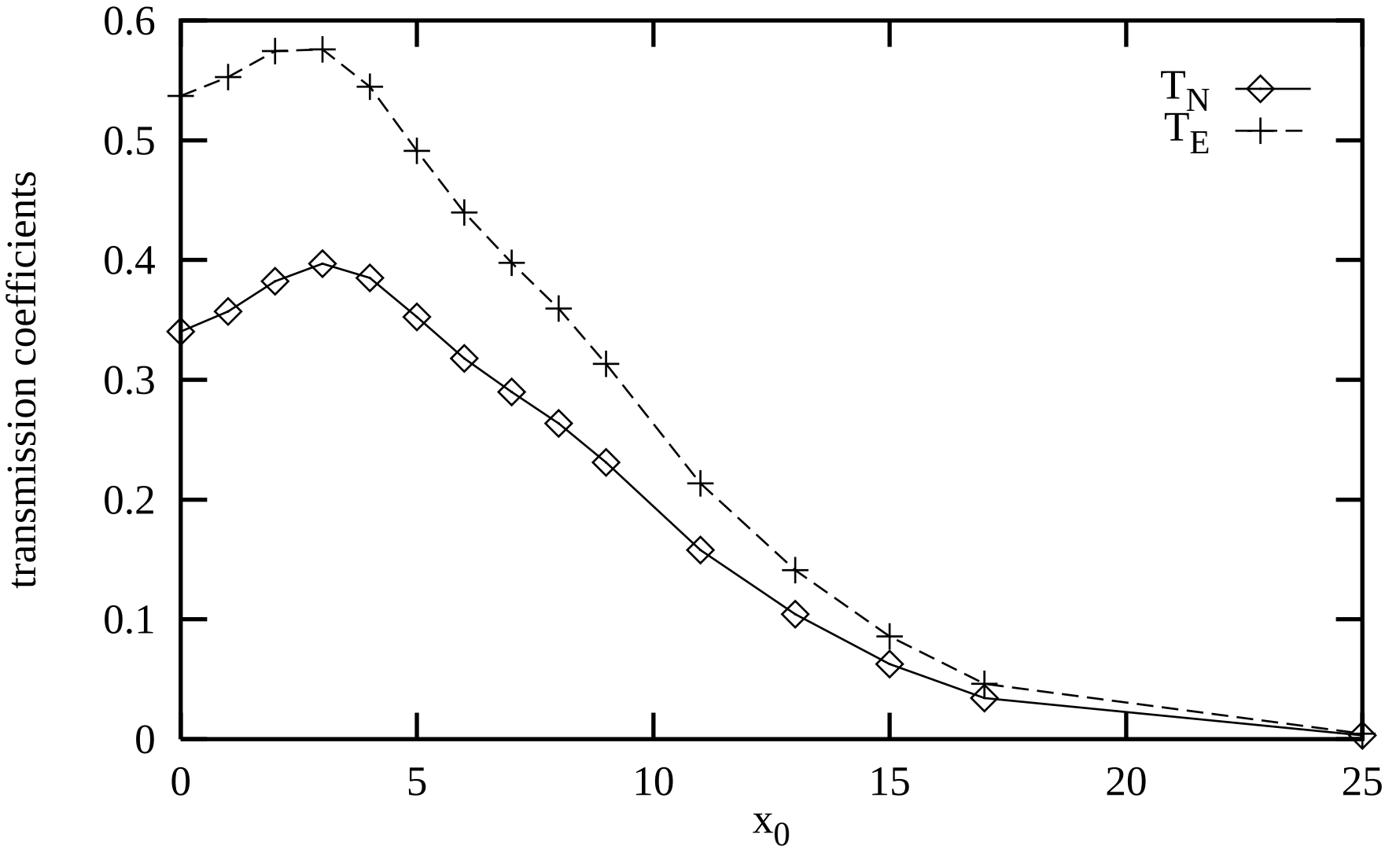

Figure 10: The transmission coefficients $T_N$ and $T_E$ for different values of $x_0$. For all curves $v_0 = 0.8$ and $V_0 = 0.35 = 1.094K_0$.